\documentclass[preprint,aps,showkeys,showpacs]{revtex4}
\usepackage{graphicx}% Include figure files
\usepackage{dcolumn}
\usepackage{bm}
\preprint{}
\def\para{\par\noindent}
\def\sqr#1#2{{\vcenter{\vbox{\hrule height.#2pt
        \hbox{\vrule width.#2pt height#1pt \kern#1pt
          \vrule width.#2pt}
        \hrule height.#2pt}}}}

\newcount\notenumber

\def\note{\advance\notenumber by 1
\footnote{$^{\the\notenumber}$}} \baselineskip 20pt
\begin{document}
\title{Persistence in a Random Bond Ising Model of Socio-Econo Dynamics}
 \author{S. \surname{Jain}}%\footnote{author for correspondence}
\email{S.Jain@aston.ac.uk}
 \affiliation{Information Engineering, The Neural
Computing Research Group, School of Engineering and Applied Science,
Aston University, Birmingham B4 7ET, U.K.}
\author{T. \surname{Yamano}}
\email{yamano@amy.hi-ho.ne.jp}
\affiliation{Department of Physics, Ochanomizu University, 2-1-1 Otsuka,
Bunkyo-ku Tokyo 112-8610, Japan}
\date[]{Received September 07 2007}

\begin{abstract}

We study the persistence phenomenon in a socio-econo dynamics model
using computer simulations at a finite temperature on hypercubic
lattices in dimensions up to 5. The model includes a \lq social\rq\
local field which contains the magnetization at time $t$. The
nearest neighbour quenched interactions are drawn from a binary
distribution which is a function of the bond concentration, $p$. The
decay of the persistence probability in the model depends on both
the spatial dimension and $p$. We find no evidence of \lq
blocking\rq\ in this model. We also discuss the implications of our
results for possible applications in the social and economic fields.
It is suggested that the absence, or otherwise, of blocking could
be used as a criterion to decide on the validity of a given model in different
scenarios.

\end{abstract}

\maketitle
\section{Introduction}
\para The persistence problem is concerned
 with the fraction of space
which persists in its initial $(t=0)$ state up to some later time
$t$. The problem has been extensively studied over the past decade
for pure spin systems at both zero [1-4] and non-zero [5]
temperatures.
\para Typically, in the non-equilibrium dynamics of spin systems
 at zero-temperature,
 the system is prepared initially in a random state and the fraction of
spins, $P(t)$, that persists in the same state as at $t=0$ up to some
later time $t$ is studied. For the pure ferromagnetic Ising model on
a square lattice the persistence probability has been found to decay
algebraically [1-4]
$$ P(t)\sim t^{-\theta},\eqno(1)$$
where $\theta\sim 0.22$ is the non-trivial persistence exponent
[1-3]. Derrida et al [4] have shown analytically that for the pure
1d Ising model $\theta = 3/8$. The actual value of $\theta$ depends on both the spin [6] and
spatial [3] dimensionalities; see Ray [7] for a recent review.
\para At criticality [5], consideration of the global order
parameter leads to a value of $\theta_{global}\sim 0.5 $ for the pure
two-dimensional Ising model.
\para It has been only fairly recently established that systems
containing disorder [8-10] exhibit different persistence behaviour
to that of pure systems. A key finding [8-9,11] is the appearance of
\lq blocking\rq\ regardless of the amount of disorder present in the
system. \lq Blocked\rq\ spins are effectively isolated from the behaviour
of the rest of the system in the sense that they {\it never} flip. As a result,
$P(\infty)>0$ and the key quantity of interest is the residual persistence given by
$$r(t) = P(t) - P(\infty).\eqno(2) $$
Note that for the five dimensional pure Ising model without any disorder blocking has also been
observed at $T=0$ [3]. At finite temperature there is no evidence of blocking [12].
\para As well as theoretical models, the persistence phenomenon
has also been studied in a wide range of experimental systems and
the value of $\theta$ ranges from $0.19$ to $1.02$ [13-15],
depending on the system. A considerable amount of the recent
theoretical effort has gone into obtaining numerical estimates of
$\theta$ for different models [1-11]. Recently, it has been found
that the behaviour of the random Ising
ferromagnet at zero temperature on a Voronoi-Delaunay lattice [16] is very similar to the
behaviour on the diluted
ferromagnetic square lattice [8,9].

\para In this work we add to the knowledge and understanding regarding
persistence by presenting the simulation results for the
behaviour of a recently proposed spin model
which appears to reproduce the intermittency observed in real
financial markets [17]. In the next section we discuss the model in
detail. In Section III we give an outline of the method
used and the values of the various parameters employed. Section IV
describes the results and the consequent implications for using the
model in a financial or social context. Finally, in Section V there
is a brief conclusion.
\section{The Model}
\para Yamano [16] has proposed a \lq minimalist\rq\ version of the Bornholdt model [18]. We
study the persistence phenomenon in the former.
In this model one has $N$ market traders, denoted by Ising
spins $S_i(t), i=1\dots
N$, located on the sites of a hypercubic lattice, $L^d=N$. The action of the
$i^{th}$ trader of buying or selling a share of a traded stock or commodity at time step $t$ corresponds to
the spin variable $S_i(t)$ assuming the value $+1$ or $-1$,
respectively. Hence, at each time step, a given trader will be either buying
or selling. A local field, $h_i(t)$, determines the dynamics of
the spins and, hence, the action of the trader. We follow [16] and assume that
$$h_i(t)=\sum_{j=1}^{2d}J_{ij}S_j(t)-\alpha\mid{\sum_{j=1}^NS_j(t)/N}\mid ,\eqno(3)$$
where the first summation runs over the nearest neighbours of $i$ only, $\alpha>0$
is a parameter coupling to the absolute magnetization. The nearest neighbour interactions
are selected randomly from
$$P(J_{ij}) = (1-p)\delta (J_{ij} +1) + p\delta (J_{ij} -1),\eqno(4)$$
where $p$ is the concentration of ferromagnetic bonds. The case $p=1/2$ corresponds to the
$\pm J$ Edwards-Anderson spin-glass [11].
Each agent is updated according
to the following heat bath dynamics:
\par
\[S_i(t+1)= \left\{
\begin{array}{c c}
+1 & \quad \mbox{with $q=[1+\exp{(-2h_i(t)/T)}]^{-1}$,}\\
-1 & \quad \mbox{with $1-q$,}\\ \end{array} \right. (5)\] where $q$
is the probability of updating and $T$ is
temperature. The first term on the right hand side in equation (3) contains the influence of the
neighbours and the second term reflects the external environment. The balance between the two terms
determines whether an agent buys or sells. If $h_i(t)=0$, the agent is equally likely to buy or sell
as $q=1/2$. If, however, $h_i(t)>0$, then $q>1/2$ and agent $i$ is more likely to buy than sell. Similarly,
if $h_i(t)<0$, then we have $q<1/2$ and the agent is more likely to sell than buy.
$\alpha$ and $T$ are tunable parameters in our model. The values we
select for these are determined by the requirement that the model should be able to
reproduce, at least qualitatively, some aspect of actual behaviour observed in a
real market. In this model the return is defined in terms of the
logarithm of the absolute value of the magnetization,
$M(t)=\sum_{j=1}^NS_j(t)/N$, that is
$$ {\rm Return}\ (t) = \ln\mid {M(t)}\mid -\ln\mid {M(t-1)}\mid\eqno(6)$$
A key stylised fact observed in real financial markets is the intermittent
or \lq bursty\rq\ behaviour in the returns [19].
Simulations [16] in spatial dimensions ranging from $d=1$ to $d=3$
confirm that the above model is able to reproduce the
required intermittent behaviour in the returns; see, for example, figure 1 in Yamano [16].
It should be emphasised that intermittency is only observed for certain values of the tunable
parameters, $\alpha$ and $T$. The temperature, $T_{int}$, is defined as the temperature at which
intermittency is observed in the returns.
Although $\alpha = 4.0$ for all dimensions considered, $T_{int}$ depending on both $d$ and
$L$. The values of $T_{int}$ as determined in [16] were: $ d=1, T_{int}=3.45, L=10001;
d=2, T_{int}=2.9, L=101; d=3, T_{int}=2.3, L=21$. A couple of points should be
emphasised at this stage. Simulations are performed only on hypercubic lattices
where the linear dimension $L$ is an odd value. This requirement ensures that the
right hand side of equation (6) is well defined. The simulations discussed in the present work were
performed in spatial dimensions ranging from $d=1$ to $d=5$. For each value of $d$, we
fine tune the parameters $\alpha$ and $T_{int}$ so that intermittent behaviour is observed
in the returns as discussed above. We found that although $\alpha =4.0$ in all cases, the
value of $T_{int}$ depends on $d$ (and also on $L$) at which intermittency is observed. Our
values for $L$ and $T_{int}$ are listed
in Table 1. Note that the values of $T_{int}$ for $d=1,2,3$ are consistent with those in
[16] but not identical because of finite-size effects.
\para In our interpretation of the model,
a persistent trader is one who doesn\rq t change his action during the course of the simulation.
 Hence, we are
interested in studying the fraction of traders who have been at
time $t$ either buying or selling continuously since $t=0$. Later, we will
also suggest a possible interpretation within the context of sociophysics of the
model.

\section{Methodology}
\para
\begin{table}
\centering
\begin{tabular}{|c|c|c|c|}
  \hline
  {Dimension} & {$L$} & {$T_{int}$} \\ \hline
   1 & 4000001 & 3.5  \\ \hline
   2 & 2001 & 3.0 \\ \hline
   3 & 151 & 2.5 \\ \hline
   4 & 45 & 1.9  \\ \hline
   5 & 21 & 1.4  \\
  \hline
\end{tabular}
\caption{Values of the linear dimension $L$ of the lattices used in
the simulations. The coupling parameter $\alpha= 4.0$ in all cases.
Intermittent behaviour was observed in the returns when the
temperature was set at $T_{int}$ as given above.} \label{datavalues}
\end{table}

\para As mentioned in the previous section, for each spatial dimension $d$ we first
fine tune the temperature to reproduce intermittent behaviour in the
returns. As can be seen from Table 1, the temperature $T_{int}(d)$
decreases with $d$. For a given dimension, all subsequent
simulations are performed at that temperature. Averages over at
least 100 samples for each run were performed and the error-bars in
the following plots are smaller than the data points.

The value of each agent at $t=0$ is noted and the dynamics updated
according to equation (3).
 \para At each time step, we count the number of agents that still
persist in their initial $(t=0)$ state by evaluating
 $$n_i(t)=(S_i(t)S_i(0)+1)/2.\eqno(7)$$
 Initially, $n_i(0)=1$ for all $i$. It changes to zero when an agent
 changes from buying to selling or vice vera for the first time.
 Note that
 once $n_i(t) =0$, it remains so for all subsequent calculations.
\para The total number, $n(t)$,
of agents who have not changed their action by time $t$ is then
given by
$$ n(t)=\sum_{i}n_i(t).\eqno(8)$$
A fundamental quantity of interest is $P(t)$, the persistence
probability. In this problem we can identify $P(t)$ with the density
of agents continuously buying or selling without interruption since the
start [1],
$$P(t)=n(t)/N,\eqno(9)$$
where $N = L^d$ is the total number of agents present.
\section{Results}
\para  We now discuss our results. We restrict ourselves to $0\le p\le 0.5$ as the
problem is symmetric about $p=0.5$. It should be noted that we tried various different fits
(exponential, power-law, stretched-exponential, etc.) for our data. We will not discuss the fits
we discarded as unsatisfactory.
\par In figure 1 we show a semi-log plot of the persistence probability
against time $t$ for a range of bond concentrations $0<p\le 0.5$ for
$d=1$. It\rq s clear from the plot that the data can be fitted to
$$P(t)\sim e^{-\gamma (p) t},\eqno(10)$$
where we estimate $\gamma (p)\sim 0.56$ from the linear fit. Note that $\gamma (p)\approx\gamma$, independently
of $p$.
\begin{figure}
\includegraphics{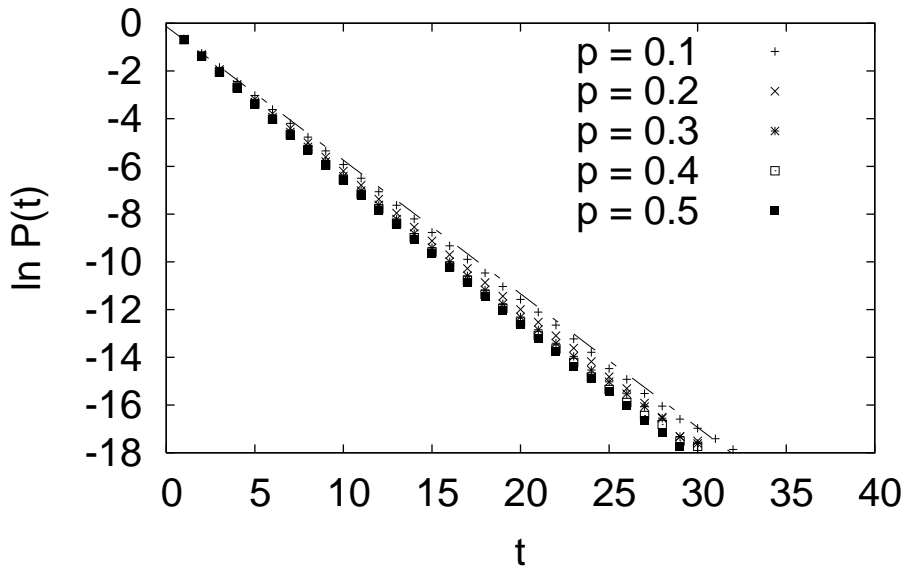}
\caption{Here we plot $\ln P(t)$ versus $t$ for $d=1$ over the range
$0.1\le p\le 0.5$. The straight line, which is a guide to the eye,
has a slope of $-0.56$.} \label{fig1}
\end{figure}
\para
Figure 2 displays the results for $d=2$. Although, just as for $d=1$, there
is evidence for exponential decay, this time it would appear that
the value of the parameter $\gamma (p)$ depends on $p$. For $p=0.1$
we estimate $\gamma (p=0.1)\sim 0.35$.
\begin{figure}
\includegraphics{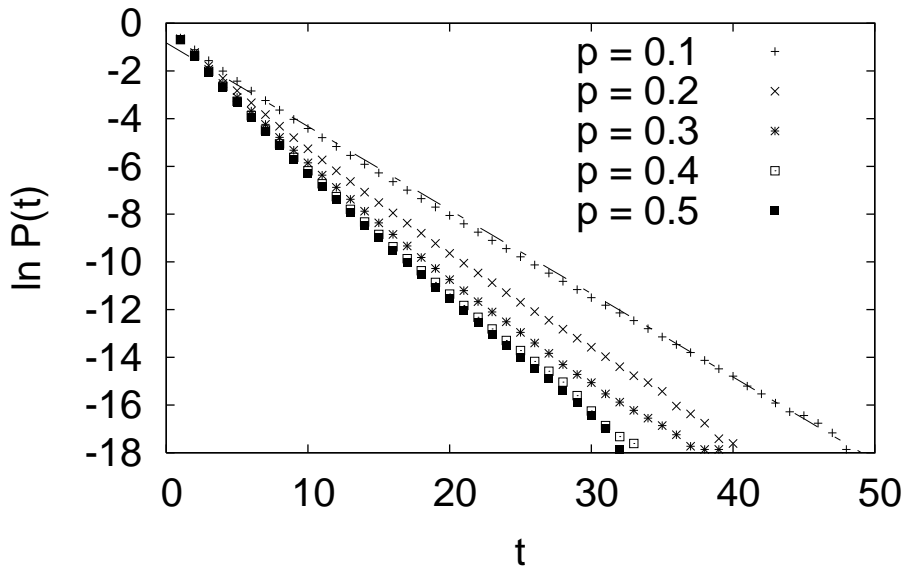}
\caption{A semi-log plot of the data for $d=2$. We see that here, in
contrast to figure 1 for $d=1$, the slopes are dependent on the bond
concentrations. The linear fit shown is that for $p=0.1$ and the
slope is $-0.35$.} \label{fig2}
\end{figure}
The results for the three-dimensional case are shown in figure 3.
Here we see clear evidence that even the qualitative nature of the decay
depends on the bond concentration. For $p=0.5$ we have behaviour
very similar to the two cases considered earlier, namely exponential
decay. The decay is clearly non-exponential for $p=0.1$ over the interval
considered. A fit to the stretched-exponential is also unsatisfactory even
though we have an additional adjustable parameter.
\begin{figure}
\includegraphics{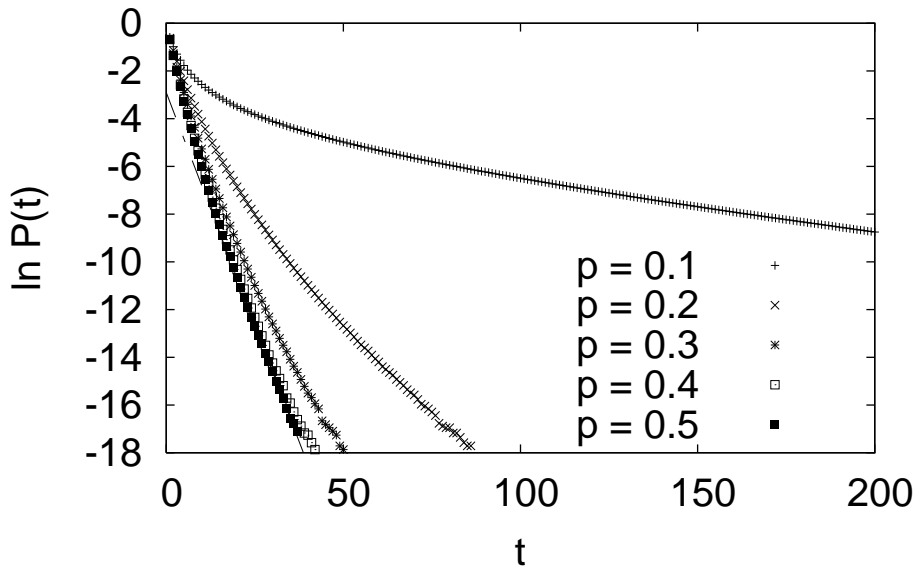}
\caption{A plot of $\ln P(t)$ against $t$ for $d=3$ for the same
bond concentrations as earlier. The straight line, which is a guide
to the eye, has a slope of $-0.39$ and indicates that the decay for
$p=0.5$ is very similar to that found in lower dimensions. The
behaviour for $p=0.1$ is clearly non-exponential.} \label{fig3}
\end{figure}
The results in $d=4$ are very similar to those for $d=3$ and we will
not present them here. Instead, in figure 4 we show a log-log plot
of the persistence against time for $d=5$. The decay of $P(t)$ is
seen to be heavily dependent on the concentration of ferromagnetic
bonds. For low values of $p (\le 0.3$), we have a power-law decay at
long times as given by equation (1) with an estimated value of
$\theta\sim 0.5$. For higher value of $p$ the decay would appear not
to be a power-law but also not exponential in it\rq s nature.
We note that even though we are working with a model containing disorder,
no \lq blocking\rq\ is observed in the simulations. This is probably because
we are working at a finite temperature and is in agreement with the earlier work
on the pure Ising model in high dimensions [3,12].
\para
\begin{figure}
\includegraphics{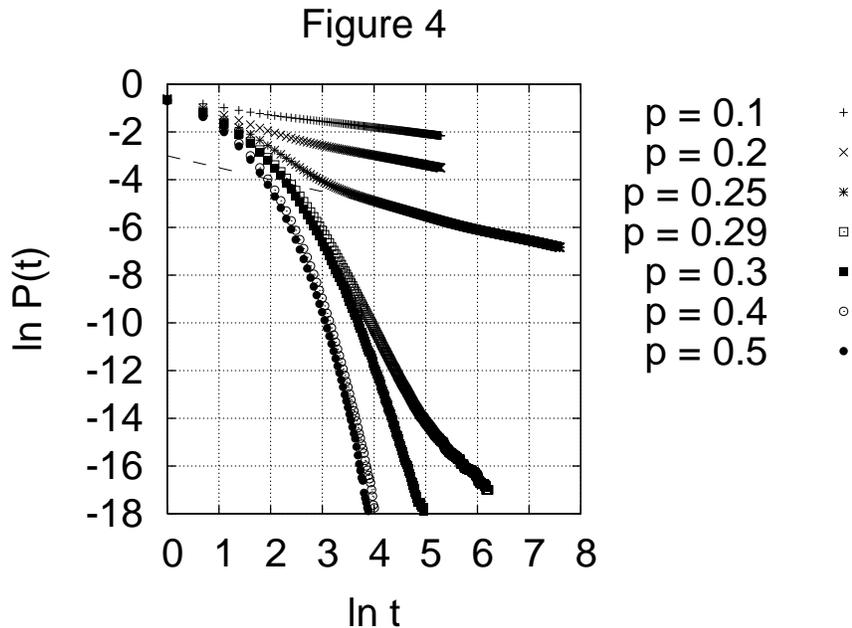}
\caption{Here we display the data for $d=5$ and selected bond
concentrations as a log-log plot. Clearly, the behaviour depends
crucially on the value of $p$. For low ($p\le 0.3$) values the decay
is power-law. The straight line shown has a slope of $-0.5$.}
\label{fig4}
\end{figure}
\para
\section{Conclusion}
\para To conclude, we have presented the results of extensive simulations
for the persistence behaviour of agents in a model capturing some of
the features found in real financial markets. Although the model
contains bond disorder, we do not find any evidence of \lq
blocking\rq\ . This is believed to be because of thermal fluctuations.
The persistence behaviour appears to depend on both
the spatial dimensionality and the concentration of ferromagnetic
bonds. Generally, whereas in low dimensions the decay is
exponential, for higher dimensions and low values of $p$ we get
power-law behaviour.

The initial model was developed in an economic context. Power law
persistence in this case means the existence of traders who keep on
buying or selling for long durations. Furthermore, the presence of
\lq blocking\rq\ would be highly unrealistic for modelling the
dynamics because the traders would have access to only a finite amount of
capital. Hence, the absence of blocking would suggest that our model is not
an unreasonable starting point for further development.

One can also interpret the model in a social context. Here the value
$S_i(t)=+1$ or $-1$ could represent an opinion. Here \lq
blocking\rq\ would be realistic and correspond to the proportion of
the population that is stubborn and not susceptible to a change. Hence, any model exhibiting
exponential decay in the persistence probability would probably be
an unrealistic model to use in this scenario.

Hence, we can use the behaviour of the persistence probability as a
criterion to decide whether we have a realistic economic or social
model.

\begin{acknowledgments}
We thank the referee for his comments on the manuscript and for
bringing [12] to our attention. TY thanks L. Pichl for allowing him
to use his CPU (mona) at International Christian University, Japan,
and acknowledges support by JSPS Grant-in-Aid \# 06225. A research
grant from Daiwa Foundation Small Grant (Ref: 6124/6356) enabled TY
to work as a non-Asian researcher in Birmingham (UK) for six days.
\end{acknowledgments}

\section*{References}
\begin{description}
\item {[1]} B. Derrida, A. J. Bray and C. Godreche, J. Phys. A:
Math Gen {\bf 27},
 L357 (1994).
\item {[2]} A. J. Bray, B. Derrida and C. Godreche, Europhys. Lett.
{\bf 27},
 177 (1994).
\item {[3]} D. Stauffer, J. Phys. A: Math Gen {\bf 27}, 5029 (1994).
\item {[4]} B. Derrida, V. Hakim and V. Pasquier, Phys. Rev. Lett.
{\bf 75},
 751 (1995); J. Stat. Phys. {\bf 85}, 763 (1996).
 \item {[5]} S. N. Majumdar, A. J. Bray, S. J. Cornell, C. Sire,  Phys.
Rev. Lett. {\bf 77}, 3704 (1996).
 \item {[6]} B. Derrida, P. M. C. de Oliveira and D. Stauffer, Physica
{\bf 224A}, 604 (1996).
 \item {[7]} P. Ray, Phase Transitions {\bf 77} (5-7), 563 (2004).
\item {[8]} S. Jain, Phys. Rev. E{\bf 59}, R2493 (1999).
\item {[9]} S. Jain, Phys. Rev. E{\bf 60}, R2445 (1999).
\item {[10]} P. Sen and S. Dasgupta, J. Phys. A: Math Gen {\bf 37},
11949 (2004)
\item {[11]} S. Jain and H. Flynn, Phys. Rev. E{\bf 73}, R025701 (2006)
\item {[12]} D. Stauffer, Int. J. Mod. Phys. C {\bf 8}, 361 (1997).
\item {[13]} B. Yurke, A. N. Pargellis, S. N. Majumdar
and C. Sire, Phys. Rev. E{\bf 56}, R40 (1997).
\item {[14]} W. Y.
Tam, R. Zeitak, K. Y. Szeto and J. Stavans, Phys. Rev. Lett. {\bf
78}, 1588 (1997).
\item {[15]} M. Marcos-Martin, D. Beysens, J-P
Bouchaud, C. Godreche and I. Yekutieli, Physica {\bf 214D}, 396
(1995).
\item {[16]} F.W.S. Lima, R.N. Costa Filho and U.M.S. Costa, Journal of Magnetism and Magnetic
Materials {\bf 270}, 182 (2004).
\item {[17]} T. Yamano, Int. J. Mod. Phys. C {\bf 13}, 645 (2002).
\item {[18]} S. Bornholdt, Int. J. Mod. Phys. C {\bf 12}, 667 (2001)
\item {[19]} R.N. Mantegna, H.E. Stanley, An Introduction to Econophysics: Correlations and Complexity
in Finance, Cambridge University Press, Cambridge (2000).
\end{description}
\end{document}